\documentclass[amsmath,nofootinbib,twocolumn]{revtex4}

\usepackage{graphicx}
\usepackage{dcolumn}
\usepackage{bm}

\newcommand*{\no}{\noindent}
\newcommand*{\bea}{\begin{eqnarray}}
\newcommand*{\eea}{\end{eqnarray}}
\newcommand*{\be}{\begin{equation}}
\newcommand*{\ee}{\end{equation}}
\newcommand*{\pd}{\partial}
\newcommand*{\pdm}{\pd_{\mu}}

\newcommand*{\pref}[1]{(\ref{#1})}

\newcommand*{\mn}{{\mu\nu}}

\newcommand*{\nn}{\nonumber}

\newcommand*{\tr}{\mathrm{tr}}

\begin{document}


\title{Accessing directly the properties of fundamental scalars in the confinement and Higgs phase}

\author{Axel Maas\footnote{Present address: Institute for Theoretical Physics, Friedrich-Schiller-University Jena, Max-Wien-Platz 1, D-07743 Jena, Germany}}

\email[]{axelmaas@web.de}
\affiliation{Institute of Physics, Karl-Franzens-University Graz,
            Universit\"atsplatz 5, A-8010 Graz, Austria}



\date{\today}

\begin{abstract}
The properties of elementary particles are encoded in their respective propagators and interaction vertices. For a SU(2) gauge theory coupled to a doublet of fundamental complex scalars these propagators are determined in both the Higgs phase and the confinement phase and compared to the Yang-Mills case, using lattice gauge theory. Since the propagators are gauge-dependent, this is done in the Landau limit of the 't Hooft gauge, permitting to also determine the ghost propagator.
It is found that neither the gauge boson nor the scalar differ qualitatively in the different cases. In particular, the gauge boson acquires a screening mass, and the scalar's screening mass is larger than the renormalized mass. Only the ghost propagator shows a significant change. Furthermore, indications are found that the consequences of the residual non-perturbative gauge freedom due to Gribov copies could be different in the confinement and the Higgs phase.
\end{abstract}
\pacs{11.15.Ex 11.15.Ha 12.15.-y 12.38.Aw 14.80.Bn}
\maketitle

\section{Introduction}

Although the theory of weak isospin can be very well described with conventional perturbation theory after pa\-ra\-me\-triz\-ing the effective Higgs condensate \cite{Bohm:2001yx}, it still poses quite a number of genuine non-perturbative questions. The most obvious one is, what happens if the Higgs is very heavy. In this case, perturbative calculations will break down eventually when reaching the energy domain of about 1 TeV \cite{Bohm:2001yx}. But this is by far not the only question. A possibly even more interesting question is, whether the theory in itself can provide a dynamical mechanism for Higgs condensation, as indicated by lattice simulations \cite{Langguth:1985dr,Jersak:1985nf,Evertz:1985fc,Langguth:1985eu,Bonati:2009pf,Fodor:2007fn}. This would obliterate the necessity to drive electroweak symmetry breaking entirely by physics beyond the standard model, though not resolving the perceived hierarchy problem \cite{Morrissey:2009tf}. Another serious obstacle is the possibility that the theory of weak isospin could indeed become trivial upon quantization, very much like the ungauged $\phi^4$ theory \cite{Callaway:1988ya} or the ungauged Higgs-Yukawa theory \cite{Fodor:2007fn,Gerhold:2007gx}. However, the possibility of a non-perturbative stabilization (similar in concept, e.\ g., to asymptotic safety) has not been finally ruled out, and deserves further investigation. Of course, it could be that the top-Yukawa coupling will be important in case of the standard model \cite{Callaway:1988ya}, but before requiring outside assistance, it should be better understood whether this is indeed necessary. Results concerning this question are not yet fully satisfactory. In particular, besides the phase diagram \cite{Jersak:1985nf,Langguth:1985eu,Bonati:2009pf} usually only gauge-invariant bound states have been investigated \cite{Evertz:1985fc,Langguth:1985dr} for this question, for which it may be difficult to separate genuine bound states from almost scattering states in a very weakly interacting theory.

Aside from these questions, which concern the practical viability of the weak isospin theory in its standard model form, there is also a more conceptual question. It has been shown that the confinement phase and the Higgs phase of the SU(2)-Higgs theory are not separated, if the theory is investigated with a (lattice-)cutoff, but are analytically connected \cite{Fradkin:1978dv}. Only in terms of gauge-dependent quantities can a phase transition be established, but the transition itself then turns out to be gauge-dependent \cite{Caudy:2007sf}. This would imply that there is no qualitative difference between a confining and a Higgs phase in terms of gauge-invariant physics. Since a confining phase is not perturbatively accessible, so would neither be a Higgs phase. This must be sorted out to get a full understanding of the field theoretical setup of the weak isospin sector of the standard model. Furthermore, latest results indicate that the thermodynamic limit for questions concerning the phase diagram is rather hard to reach \cite{Bonati:2009pf}.

On the other hand, this also permits to study an entirely different set of problems in the same theory. In the confinement phase, the color confinement process appears to be entirely the same as for quarks, but without the complications introduced due to chiral symmetry and fermions\footnote{Note that the often cited Wilson criterion for confinement \cite{Greensite:2003bk} is actually blind to the spin structure of the involved particles.}. For an understanding of color confinement, the weak isospin theory in the confinement phase therefore offers an ideal laboratory, as has already been exploited previously \cite{Knechtli:1998gf,Knechtli:1999qe,Philipsen:1998de,Fister:2010yw,Fischer:2009tn}. This is another incentive to study this theory in both phases.

None of the points above, with the possible exception of a heavy Higgs, imply that for precision measurements in the standard model non-perturbative physics is necessary. Non-perturbative effects could easily be sufficiently suppressed over the whole accessible energy spectrum, even at LHC, or possibly up to the Planck mass, to make these questions practically irrelevant. And in fact, the results from LEP and Tevatron so far are in perfect agreement with a domination of perturbative contributions. However, as long as it is not clear what happens at LHC and beyond, the consequences of non-perturbative contributions should be known.

To find answers to the questions above, in this work the possibility to access the properties of the (gauge-de\-pen\-dent) elementary degrees of freedom of the weak isospin theory will be investigated. These are the gauge bosons, which will be called the $W$ for simplicity. Of course, since QED is neglected, the $Z$ and $W$ are degenerated here. The second will be denoted Higgs. Their propagators will be determined in both the Higgs and confinement phase, and will be compared to the quenched case, i.\ e.\ without dynamical Higgs particles, and thus a Yang-Mills theory with gauge group SU(2). For the non-perturbative determination a lattice implementation will be used.

Since the propagators are gauge-dependent, it is necessary to fix a gauge. Given that non-perturbative gauge-fixing is made complicated by the Gribov-Singer ambiguity \cite{Gribov:1977wm,Singer:1978dk}, a gauge is chosen in which these effects are comparatively well understood, which is the Landau-limit of the 't Hooft gauge, corresponding to the Landau gauge in Yang-Mills theory. Thus, as additional degrees of freedom, the Faddeev-Popov ghosts, also all degenerate, are available. As a consequence of the Gribov-Singer ambiguity, on a finite lattice, as employed here, the Landau gauge becomes a family of non-perturbative gauges in Yang-Mills theory \cite{Maas:2009se}. This effect will also be studied here.

This, rather exploratory, investigation is structured as follows: In section \ref{stech} the technical details of the simulations, including gauge-fixing, will be briefly discussed. In section \ref{sprop} the propagators and their renormalization will be introduced and the central results presented. In section \ref{sgauge} the consequences of the Gribov-Singer ambiguity will be analyzed. Everything is summarized in section \ref{ssum}. Some technical details are deferred to an appendix.

\section{Technicalities}\label{stech}

The method used here is essentially standard lattice gauge theory. The lattice version of the weak SU(2) isospin model\footnote{Note that strictly speaking the gauge group is SU(2)/Z$_2$ due to the requirement of anomaly cancellation when including fermions \cite{O'Raifeartaigh:1986vq}. Here, instead, the $Z_2$ part is merely explicitly broken, and it is assumed that this does not make a difference on the level of correlation functions.} is given by \cite{Montvay:1994cy}
\bea
S&=&\beta\sum_x\Big(1-\frac{1}{2}\sum_{\mu<\nu}\Re\tr U_\mn(x)\nn\\
&&+\frac{1}{2}\phi^+(x)\phi(x)+\lambda\left(\phi(x)^+\phi(x)-1\right)^2\nn\\
&&-\kappa\sum_\mu\Big(\phi(x)^+U_\mu(x)\phi(x+e_\mu)\nn\\
&&+\phi(x+e_\mu)^+U_\mu(x)^+\phi(x)\Big)\Big)\nn\\
U_\mn(x)&=&U_\mu(x)U_\nu(x+e_\mu)U_\mu(x+e_\nu)^+U_\nu(x)^+\nn\\
W_\mu&=&\frac{1}{2agi}(U_\mu(x)-U_\mu(x)^+)+{\cal O}(a^2)\nn\\
\beta&=&\frac{4}{g^2}\nn\\
a^2m_0^2&=&\frac{(1-2\lambda)}{\kappa}-8\nn.
\eea
\no In this expression $a$ is the lattice spacing, $W_\mu$ the gauge boson field, $\phi$ the Higgs field, $g$ the bare gauge coupling, $\lambda/\kappa^2$ is the bare self-interaction coupling of the Higgs, $m_0$ the bare mass of the Higgs, the sums are over the lattice points $x$, and $e_\mu$ are unit vectors on the lattice in the direction $\mu$.

For such a system the path integral, and thus the expectation values of operators, can be evaluated using Monte Carlo simulations. The method used for the gauge part is described in detail in \cite{Cucchieri:2006tf}. For the Higgs part, for each of the six sub-sweeps of the hybrid-overrelaxation cycle of the gauge field a local Metropolis update has been performed, where the acceptance probability was adjusted adaptively to be 50\%. The results for observables like the action, the spatial average of the expectation value of the operator
\be
\eta(x)=\phi(x)^+\phi(x)\label{higgsonium},
\ee
\no the expectation value of the plaquette, and the lowest masses of the isoscalar-scalar and isovector-vector excitations have been compared to literature values \cite{Langguth:1985dr} to check the code.

In the following, three different sets of parameters will be investigated, all for a lattice size of $N^4=24^4$. One is the quenched case, i.\ e., the Higgs field is set to zero, while the other two cases correspond to systems deep inside the confinement phase and the Higgs phase\footnote{Though these phases are not strictly separated on a finite lattice, it is possible to determine a phase diagram. Being sufficiently far away from the cross-over region, the systems are then classified by these names, though they are strictly speaking qualitatively identical. Furthermore, in the sense of the Wilson criterion, the confinement phase is not confining due to screening by pair creation \cite{Greensite:2003bk}, and in the Higgs phase the symmetry is actually not broken, but only hidden \cite{Bohm:2001yx,Caudy:2007sf}. Therefore, these terms should be taken only as a short-hand notice, and their true meaning kept in mind all the time.} \cite{Knechtli:1999qe}.

A certain problem is posed by translating the lattice scale to a physical scale. In the Higgs phase it could be suggested to match with the Higgs condensate. However, this quantity is gauge-dependent, and in particular zero in the Landau gauge \cite{Caudy:2007sf}. The possibility to instead use the spatially averaged expectation value of $\eta$ is also not useful: It is non-zero also in the confinement phase \cite{Langguth:1985dr}, and it is therefore dubious to associate it with the Higgs condensate. The next possibility would be to match the $W$ boson pole mass, which is measured in experiment. There are two reasons which make this approach rather hard. The first is that the pole positions would have to be determined by analytical continuation, which is rather unreliable with the number of lattice points available here. The second problem is that, at least in Yang-Mills theory, and likely also at least in the confinement phase, the $W$ boson does not exhibit a pole \cite{Cucchieri:2004mf,Bowman:2007du,Alkofer:2003jj,Fischer:2008uz}. Therefore, also this possibility is not useful for the present purpose. Concerning the Higgs, similar considerations apply, even if its mass would have been determined in experiment.

Hence a more ad-hoc procedure will be used here. For both the confinement and the Higgs phase, the mass $m_\eta$ of the lowest lying state of the composite operator $\eta(x)$, \pref{higgsonium}, will be determined. This state would be an isoscalar-scalar Higgsonium bound state (or scattering state). Its mass will be arbitrarily set to 250 GeV, motivated that it may be twice as heavy as a single Higgs with its most possible mass of about 125 GeV \cite{Haller:2010zb}. Since, in contrast to quarks in QCD, it is found below that the Higgs screening mass and its renormalized pole mass turn out to be rather close, it could be expected on the basis of a simple constituent model that this should give an acceptable first guess. In the quenched case, this object is rather tedious to calculate. Instead, here an upscaled version of the $a(\beta)$ relation of Yang-Mills theory  will be used, to set a scale compatible with the one in the confinement phase. That said and done, one should see the physical scale to be rather of illustrative purpose, and it can always be scaled out again to replace it with another scale.

It should be noted\footnote{I am grateful to Christian Lang for a discussion of this issue.} that in the lattice literature the operator \pref{higgsonium} is usually associated  with the Higgs itself rather than with a Higgs-Higgs-bound state \cite{Langguth:1985dr,Evertz:1985fc}, and thus its mass is denoted as the Higgs mass. The motivation for this is that it is the simplest operator available, and, provided no anomalous hierarchy is encountered, it will also be the lightest state in the isoscalar sector, and thus the lightest physical excitation of the theory. In particular, it is the square of the radial mode of the elementary Higgs field, as being its lowest order gauge-invariant contribution in an expansion in terms of polynomials in the fields. However, from the point of view of the gauge-dependent elementary fields, it is a composite bound-state, very much like the $\sigma$-meson in QCD. It is thus not directly associated with the usual perturbative definition of the Higgs \cite{Bohm:2001yx} in 't Hooft gauges where the Higgs is not an isoscalar. On the other hand, some different relation may exist in the non-renormalizable unitary gauge. Anyway, it should be noted that neither the pole mass nor the screening mass of an elementary Higgs is a renormalization-group invariant, even if it should be gauge-invariant. In contrast, the mass of the lowest excitation in the $\eta$ channel is renormalization-group-invariant. Thus a direct identification of both concepts is, at least, not obvious.

Part of the investigation here will therefore be to study how these two concepts are related\footnote{Similar considerations apply to the lowest isovector vector channel investigated in lattice calculations \cite{Evertz:1985fc,Langguth:1985dr}, which is a collective excitation of elementary Higgs and $W$ bosons, but the lowest gauge-invariant vector-particle state.}. One possible outcome could, e.\ g., be a similar relation as for quarks and mesons in QCD.

\begin{table*}
\caption{\label{config}The three sets included for comparison. For setting the scale $a$, see the text and appendix \ref{app:mass}. The number of configurations are given for the determination of the propagators, where 1080 thermalization sweeps and 108 decorrelation sweeps have been performed in multiple independent runs. In the quenched case, the first number is for the $W$ and ghost propagator, the second for the quenched Higgs propagator. $m_0$ is the tree-level mass of the Higgs. In the quenched case this is the mass appearing in the covariant Laplacian \pref{covlap}, see section \ref{sprop}.}
\vspace{1mm}
\begin{tabular}{|c|c|c|c|c|c|c|c|c|c|}
\hline
System & $am_\eta$ & $a^{-1}$ [GeV] & $\beta$ & $m_0^2$ [GeV$^2$] & $\kappa$ & $\lambda$ & Propagator\cr
\hline
Quenched & - & 114(7) & 2.2 & (123$^2$) & - & - & 95/194\cr
\hline
Confinement & $3.2(2)$ & 78(5) & 2.0 & -(221$^2$) & 0.25 & 0.5 & 623\cr
\hline
Higgs & $1.08^{+0.06}_{-0.04}$ & $231_{-9}^{+13}$ & 2.3 & -(654$^2$) & 0.32 & 1.0 & 1227\cr
\hline
\end{tabular}
\end{table*}

That completes the basic idea used here to determine the scale. The details of the determination of the scale as implemented here are deferred to appendix \ref{app:mass}, where also systematic uncertainties are discussed. The resulting set of parameters is then given in table \ref{config}. The bare couplings have been chosen such that the systems are sufficiently far away from the phase transition such that no meta-stable phases could occur while at the same time the lattice spacing is not exceedingly small or large, based on the results in \cite{Langguth:1985dr,Evertz:1985fc,Knechtli:1999qe}. 

This concludes the generation of the configurations. It remains to gauge-fix them. As stated in the introduction, this is potentially obstructed by the presence of Gribov copies. However, the presence of Gribov copies can also be turned into a virtue by using them to define different non-perturbative gauges, depending on the selection of certain Gribov copies \cite{Fischer:2008uz,Maas:2008ri,Maas:2009se}. Still, all of these non-perturbative gauges satisfy the associated perturbative gauge condition, which therefore has to be chosen first.

In the present case, this will be the Landau limit of the 't Hooft gauge. Thus, the gauge fields satisfy the condition
\be
\pdm W_\mu=0\label{lgc}
\ee
\no in the now Euclidean space time. To deal with the Gribov copies, first of all the selection of copies is restricted to the first Gribov horizon, defined to be the region with strictly positive semi-definite Faddeev-Popov operator \cite{Gribov:1977wm}.

The Gribov-Singer ambiguity now leads to the fact that potentially more than one gauge copy of a given configuration satisfies both conditions. In fact, in Yang-Mills theory appear to exist a large, possibly infinite, number of such copies in the continuum and infinite-volume limit \cite{vanBaal:1997gu,Mehta:2009zv}. At least in a finite volume, these can be used to design different gauges \cite{Maas:2008ri,Maas:2009se}. In the next section, the so-called minimal Landau gauge \cite{Cucchieri:1995pn} will be used, which selects one random representative among the remaining Gribov copies to represent a configuration. The consequences of alternative choices will be discussed in section \ref{sgauge}, a detailed description can be found in \cite{Fischer:2008uz,Maas:2008ri,Maas:2009se}.

The method used to fix this gauge is described in \cite{Cucchieri:2006tf}. Since the gauge condition only involves the gauge fields, the same method for gauge-fixing as in Yang-Mills theory can be used. After obtaining the SU(2)-valued gauge-transformation $g(x)$, the gauge-fixed Higgs field $\phi'$ is obtained as
\be
\phi'(x)=g(x)\phi(x)\nn,
\ee
\no as required to make the action gauge-invariant.

\section{Propagators}\label{sprop}

Since Landau gauge is weak-isospin symmetric, and thus the vacuum expectation value of the Higgs field vanishes \cite{Caudy:2007sf}, the isospin symmetry is fully conserved and manifest even in the Higgs phase, as a consequence of Elitzur's theorem \cite{Elitzur:1975im}. Thus, also the propagators are isospin symmetric. There are then three independent ones, the $W$ propagator, the Higgs propagator, and the Faddeev-Popov ghost \cite{Bohm:2001yx}.

\begin{figure*}
\includegraphics[width=\linewidth]{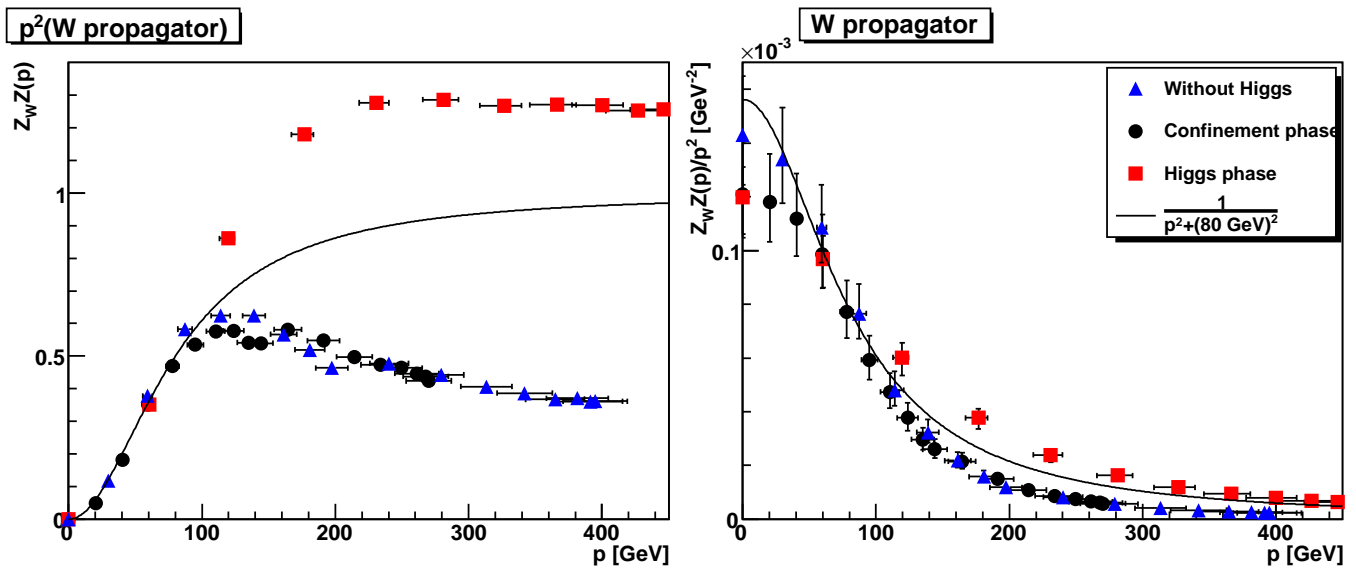}
\caption{\label{gp}The $W$ propagator (right) and dressing function (left). Momenta are along the $x$-axis for low momenta and along the $xy$-diagonal for high momenta, such that violations of rotational symmetry for the displayed points are negligible. For larger momenta along the axes they become of the order of 10-20\%. The quenched, confinement, and Higgs case are denoted by triangles, circles, and squares, respectively. Statistical errors in this section contain contributions from both statistical fluctuations and the scale uncertainty, the latter dominating. Note that the dressing function is dimensionless, and is therefore not affected by the scale uncertainty.}
\end{figure*}

The $W$ propagator in Landau gauge is given by
\be
D_\mn^{ab}=\delta^{ab}\left(\delta_\mn-\frac{p_\mu p_\nu}{p^2}\right)\frac{Z_WZ(p^2)}{p^2}\nn.
\ee
\no The methods used to determine it and the ghost propagator below can be found in \cite{Cucchieri:2006tf}. In Landau gauge, one renormalization condition is required to fix the wave function renormalization constant\footnote{For simplicity, all wave-function renormalization constants are taken to multiply the propagators, not always in line with standard conventions. However, since only the combination $Z_WZ$ etc.\ play a role here, this is essentially irrelevant.} $Z_W$. It will be conveniently chosen to satisfy
\be
Z_WZ(\mu^2)=\frac{\mu^2}{\mu^2+m_W^2}\nn,
\ee
\no with $\mu=80$ GeV and $m_W=80$ GeV. If the propagator would be at tree-level, this would yield roughly the experimentally observed pole mass of the $W$ boson. The resulting propagator $Z_WZ(p^2)/p^2$ and the dressing function $Z_WZ(p)$ are shown in figure \ref{gp}.

The results show that there is almost no difference between the case without Higgs and the confinement phase. In particular, in both cases a non-zero screening mass exists\footnote{The actual value of the screening mass is very sensitive to lattice artifacts \cite{Cucchieri:2007rg,Bogolubsky:2009dc,Sternbeck:2007ug}, and the value, which can be read from the plot, should be taken with care.}, which is about $1.1m_W$. For the Yang-Mills case, it is known that, despite its simple appearance, the analytic structure of the propagator is rather involved \cite{Cucchieri:2004mf,Bowman:2007du,Alkofer:2003jj,Fischer:2008uz}. In particular, though a non-zero screening mass is present, there appears to be no pole mass, and the particle has no representation as a K\"allen-Lehmann state. A first glimpse indicates that the corresponding Schwinger function \cite{Alkofer:2003jj} in the confinement phase has essentially the same form as in Yang-Mills theory, suggesting that this is not changing. Nonetheless, this requires further investigation, in particular for larger lattice volumes and better discretizations.

In the Higgs phase, the propagator is quantitatively different from the confinement phase. In particular, it is much closer to a tree-level behavior. Nonetheless, there are some differences, and in particular its screening mass is larger than the mass at the renormalization point, about $1.1m_W$. Therefore, there are sizable corrections to its behavior. Still, these are essentially quantitative effects\footnote{Note that the apparent absence of a maximum of the dressing function in the Higgs phase is misleading. At sufficiently large momenta propagators in both phases will coincide, and both run logarithmically to zero essentially perturbatively \cite{Bohm:2001yx}. Hence, in all cases there is a maximum, only its height and width change.}. Concerning the analytic structure, first indications hint that at least quantitatively the Schwinger function in the Higgs phase may be substantially different. However, a qualitative difference to the confinement phase cannot be discerned unambiguously without larger and finer lattices. Therefore, this is not shown explicitly here.

Instead of now directly investigating the Higgs propagator, it is interesting to investigate the ghost first. Its propagator is a scalar function given by
\be
D_G^{ab}=-\delta^{ab}\frac{Z_GG(p^2)}{p^2}\nn.
\ee
\no In the Landau gauge, its tree-level mass is zero \cite{Bohm:2001yx}, and thus the renormalization condition
\be
Z_GG(\mu^2)=1\nn,
\ee
\no will be used, with $\mu=80$ GeV. The reason why this gauge degree of freedom is interesting is that in Landau gauge it is possible to combine it with the $W$ propagator to obtain an expression for the renormalization-group invariant running gauge coupling as \cite{vonSmekal:1997is,vonSmekal:1997vx}
\be
\alpha(p^2)=\alpha(\mu^2)Z_WZ(p^2)(Z_GG(p^2))^2\label{alpha},
\ee
\no where the dependence on the renormalization scale of the propagators has been suppressed. This coupling is given in the so-called miniMOM scheme, but can be translated into the $\overline{\mathrm{MS}}$ scheme, and is known up to four loops perturbatively \cite{vonSmekal:2009ae}.

\begin{figure*}
\includegraphics[width=\linewidth]{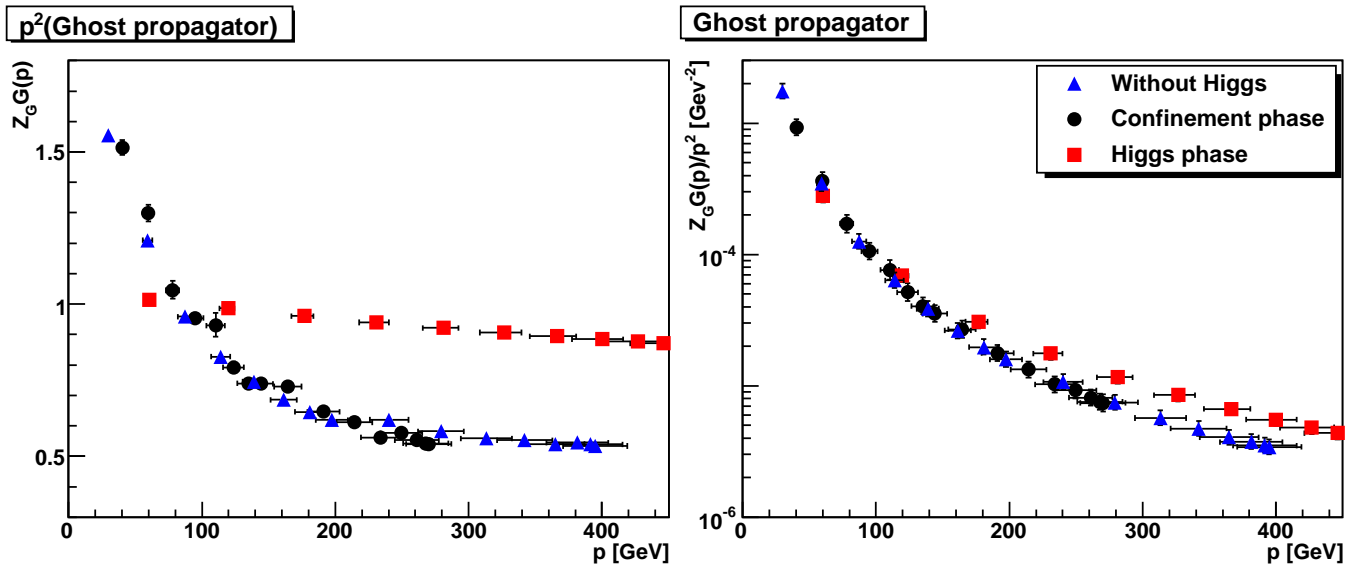}
\caption{\label{ghp}The ghost propagator (right) and dressing function (left). Momenta are along the $x$-axis for low momenta and along the $xy$-diagonal for high momenta. The quenched, confinement, and Higgs case are denoted by triangles, circles, and squares, respectively.}
\end{figure*}

The result for the ghost propagator is shown in figure \ref{ghp}. It is immediately clear that there is a drastic difference between the confinement phase, again being essentially identical to Yang-Mills theory, and the Higgs phase. The significance of the infrared enhancement seen\footnote{At very small momenta the ghost dressing actually becomes also finite or logarithmically divergent \cite{Cucchieri:2008fc}. However, the significance of this is currently still not fully resolved, see, e.\ g.\ \cite{Fischer:2008uz,Dudal:2009xh,Braun:2007bx}, for detailed discussions and in particular for further references.} is likely associated  with the confinement process. In marked contrast is the ghost propagator in the Higgs phase, where it is almost a bare propagator, without any dressing. Indirect evidence for this drastic difference has already been obtained earlier from calculations in Coulomb gauge \cite{Greensite:2004ur}, and has been discussed as a possibility in linear covariant gauges \cite{Kugo:1979gm,Alkofer:2000wg,Lerche:2002ep}.

\begin{figure}
\includegraphics[width=\linewidth]{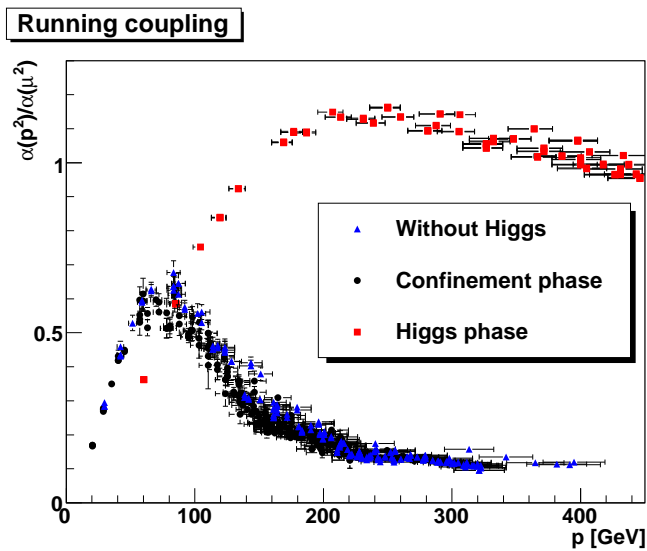}
\caption{\label{fig:alpha}The effective coupling \pref{alpha}. Momenta are along various directions \cite{Cucchieri:2006tf}. The quenched, confinement, and Higgs case are denoted by triangles, circles, and squares, respectively. Note that the overall scale is arbitrary after division by $\alpha(\mu)$.}
\end{figure}

Hence, the most distinct difference between the confinement and Higgs phase so far is the rather different ghost propagator. This also finds its manifestation in the effective coupling \pref{alpha}, as shown in figure \ref{fig:alpha}. It is visible that the coupling has no Landau pole in the confinement phase, resembling the situation in the Yang-Mills case \cite{Cucchieri:2007rg,Bogolubsky:2009dc,Sternbeck:2007ug} once more. On the other hand, the coupling is in all cases infrared suppressed in this gauge, and starts to follow the same qualitative behavior at large momenta. Thus, from this point of view the only difference between the Higgs and the confinement phase is in the size of the coupling, rather than its low-momentum behavior.

The final propagator is then the Higgs propagator. Its bare lattice version is obtained from the Fourier-trans\-for\-med Higgs field
\be
\phi'(p)=\sum_x e^{i\frac{2\pi px}{N}}\phi'(x)\nn,
\ee
\no where $p$ and $x$ are lattice coordinates and momenta, and a symmetric lattice is assumed. The lattice Higgs propagator is then given by
\be
D_H^{Lab}=\frac{\kappa}{V}<\phi'(p)^{a+}\phi'(p)^{b}>\nn,
\ee
\no where the complex conjugation also inverts the direction of the momentum, as usual. This propagator can be assigned physical units in the same way as for the $W$ propagator to obtain the continuum propagator \cite{Cucchieri:2006tf}.

In the quenched case, the elementary Higgs field $\phi$ is not available. The quenched propagator is then obtained, in analogy to the quenched quark propagator, by the inversion of an operator. For a fundamental Higgs, this is just the fundamental covariant Laplacian with a mass term,
\be
-D^2=-\left(\pdm-i\frac{g\tau^a}{2}W_\mu^a\right)^2\nn+m_0^2,
\ee
\no with the generators of the gauge algebra $\tau^a$, which on the lattice is given by
\bea
-D^2_L&=&-\sum_\mu\Big(U_\mu(x)\delta_{y(x+e_\mu)}+U_\mu^+(x-\mu)\delta_{y(x-e_\mu)}-2{\bf 1}\delta_{xy}\Big)\nn\\
&&+m_0^2{\bf 1}\delta_{xy}\label{covlap},
\eea
\no where ${\bf 1}$ is a unit matrix in weak isospin space. Since this operator is positive semi-definite, it can be inverted. For this purpose, the same method has been used as in case of the Faddeev-Popov operator in \cite{Cucchieri:2006tf}. It should be noted that even a zero mass is not a problem for this method\footnote{In contrast to the Faddeev-Popov operator, this operator has no trivial zero modes, and thus an inversion even at zero momentum is possible. However, since constant modes affect the result on a finite lattice, this is not done here.}. With this, the bare scalar propagator is available for all systems.

In contrast to the ghost and $W$ propagator the renormalization of the scalar propagator is somewhat more complicated, since besides the multiplicative wave-function renormalization also an additive mass renormalization is necessary. The renormalized propagator is given by \cite{Bohm:2001yx}
\be
D_H^{ab}(p^2)=\frac{\delta^{ab}}{Z_H(p^2+m^2)+\Pi_H(p^2)+\delta m^2}\nn,
\ee
\no where $\Pi_H$ is its self-energy, and $Z_H$ and $\delta m^2$ are the wave-function and mass renormalization constants, respectively. The two renormalization conditions implemented here are
\bea
D_H^{ab}(\mu^2)&=&\frac{\delta^{ab}}{\mu^2+m_H^2}\nn\\
\frac{\pd D_H^{ab}}{\pd p}(\mu^2)&=&-\frac{2\mu\delta^{ab}}{(\mu^2+m_H^2)^2}\nn,
\eea
\no where $\mu=125$ GeV and the would-be pole mass $m_H$ is chosen to be 125 GeV. Using a derivative with respect to the momentum instead of the momentum squared is more appropriate for the lattice with its more-or-less evenly spaced momenta at the most reliable intermediate scales than the more conventional momentum squared. Of course, this is also sufficient to determine the two unknown renormalization constants, though the result is therefore a bit different from the usually employed schemes.

\begin{figure}
\includegraphics[width=\linewidth]{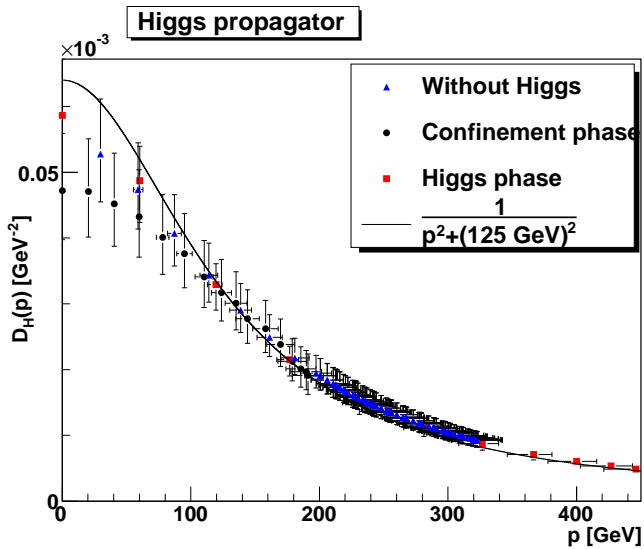}
\caption{\label{fig:sp}The Higgs propagator. Momenta are along the $x$-axis for low momenta and along the $xy$-diagonal for high momenta. The quenched, confinement, and Higgs case are denoted by triangles, circles, and squares, respectively.}
\end{figure}

The results for the renormalized Higgs propagator are shown in figure \ref{fig:sp}. It is immediately visible that the propagator in all cases deviates only slightly from an almost bare propagator with the renormalized mass. The only significant deviations from the bare propagator are found at small momenta, and the screening mass $D_H(0)^{-1/2}$ is in all cases larger than the renormalized mass. In the Higgs phase, its value is only slightly larger, about 130 GeV, while in the confinement phase it is about 145 GeV. Thus, there remain some non-trivial infrared modifications of the Higgs propagation. Other than that, the Higgs propagator is essentially unaffected compared to its tree-level behavior. Thus, even in the confinement phase, little is visible from the non-perturbative interactions.

It should be noted that the propagators in the Higgs phase all exhibit the perturbatively expected behavior. In particular, a non-zero screening mass is found for the $W$ and the negative mass squared of the Higgs became a positive effective mass. Thus, despite the vanishing vacuum expectation value and thus the not explicitly hidden symmetry, the dynamics seem to be Higgs-like, and thus to be dynamically generated. However, one should be wary about this statement for two reasons: First, this is a result on a single lattice, and lattice artifacts may play a significant role. Secondly, this is not qualitatively different from the confinement phase, again reemphasizing that the non-perturbative distinction of the Higgs and the confinement phase is still rather obscure.

Finally, both the renormalized mass and the screening mass of the elementary Higgs agree with a very simple-minded constituent Higgs model for the Higgsonium state \pref{higgsonium}. This therefore supports the view of the elementary Higgs building up the gauge-invariant states in the form of bound-states.

\section{Non-perturbative gauge-dependence}\label{sgauge}

As noted, the Gribov-Singer ambiguity implies that in principle there exists more than one solution to the gauge condition \pref{lgc}. The gauge used so far reduces the number of Gribov copies by fixing to the first Gribov region in Landau gauge, and then choosing randomly a representative from the residual gauge orbit left. This implies that in this minimal Landau gauge the obtained gauge-dependent correlation functions are effectively averaged with a certain weight over the residual gauge orbits \cite{Maas:2009se}. To the best of our current knowledge, this weight function is flat, and thus the distribution on the gauge orbits is faithfully reproduced.

This is only one admissible way to treat the residual gauge freedom. Many alternatives have been investigated, see e.\ g.\ \cite{Maas:2009se,Maas:2008ri,Bornyakov:2008yx,Cucchieri:1997dx,Silva:2004bv} . In Yang-Mills theory, at least at finite volumes and discretizations the correlation functions depend on this non-perturbative choice, in general up to the typical scale $\Lambda_\mathrm{YM}$. Though of course this dependency is irrelevant for the determination of gauge-invariant physical observables like cross-sections, it can be useful to choose an alternate gauge for technical reasons. E.\ g., there exists hints that some non-perturbative gauges in Yang-Mills theory may be more amendable for the explicit construction of the Hilbert space, while others, due to a lack of infrared singularities, are more useful in various practical calculations. It is therefore worthwhile to study the dependence of correlation functions on this gauge choice also in the case with matter fields. 

For that purpose, the methods described in \cite{Maas:2009se,Cucchieri:1997dx} will be adopted here. In particular, Gribov copies are then searched for by a multi-start algorithm in the gauge-fixing procedure. In the present case, five random starts are performed for each configuration, since only an indicative result is desired here. It is inherent to this method that only a lower limit to the gauge dependence of the propagators can be achieved.

The first finding is that no Gribov copies are found in the Higgs phase. Besides the interesting option that indeed in the Higgs phase only configurations contribute to the path integral significantly which have no Gribov copies inside the first Gribov region, the second possibility is that the strongly volume and discretization-dependent \cite{Maas:2009se} number of Gribov copies for the selected parameters is just so small that none have been found. Given the experience with results on small two-dimensional volumes in Yang-Mills theory \cite{Maas:2008ri,Maas:2009se}, where also very few copies are present, and the appreciable variation of the propagators in the investigated momentum window here, either possibility would indicate that the residual gauge orbit in the Higgs phase has a significantly different structure than in the confinement phase.

An alternative would be that the employed multi-start algorithm is just not successful in finding copies. Again, given its successes for the Yang-Mills theory \cite{Maas:2009se,Cucchieri:1997dx,Silva:2004bv}, this would imply a significantly different structure of the residual gauge orbit in the Higgs phase.

Irrespective of the precise reason the residual gauge dependence of the correlation functions in the Higgs phase is thus not existent, and in the following only the confinement phase will be investigated, and compared to the Yang-Mills case. In that case, the average number of Gribov copies found almost saturates the number of Gribov copies checked, i.\ e., there have been found 4.38(3) copies in the confinement phase per configuration in five attempts. In the Yang-Mills case, this number is slightly larger, 4.55(6).

To provide an estimate of the variability of the correlation functions, these will be determined for the absolute Landau gauge and the max-$B$ gauge, see \cite{Maas:2009se} for their definition. These gauges have been found so far to limit the variability of the correlation functions in Yang-Mills theory at a fixed number of Gribov copies. The first of these gauges attempts to minimize the $W$ propagator by the choice of the Gribov copy, while the second attempts to maximize the ghost propagator in the infrared.

\begin{figure*}
\includegraphics[width=0.5\linewidth]{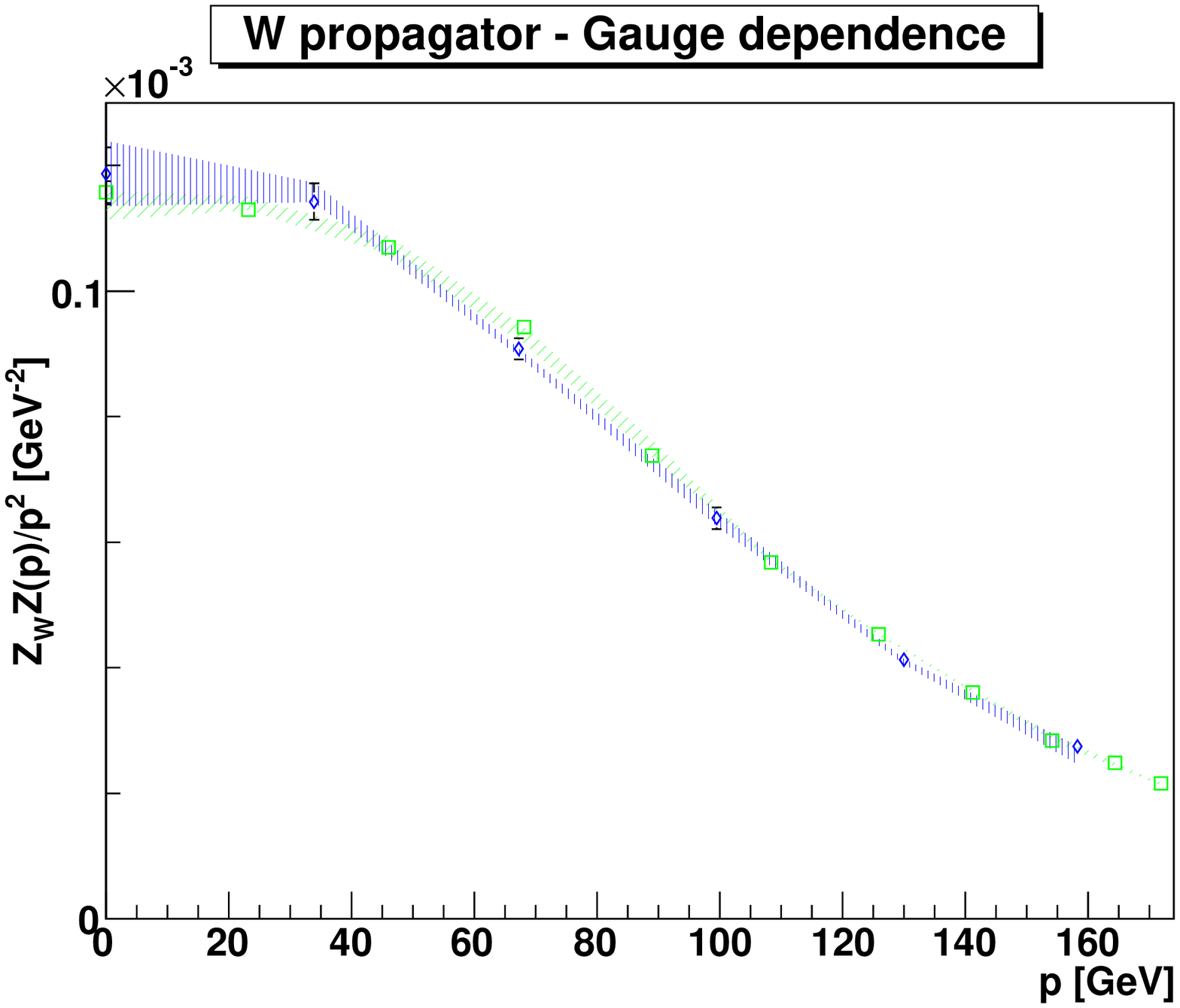}\includegraphics[width=0.5\linewidth]{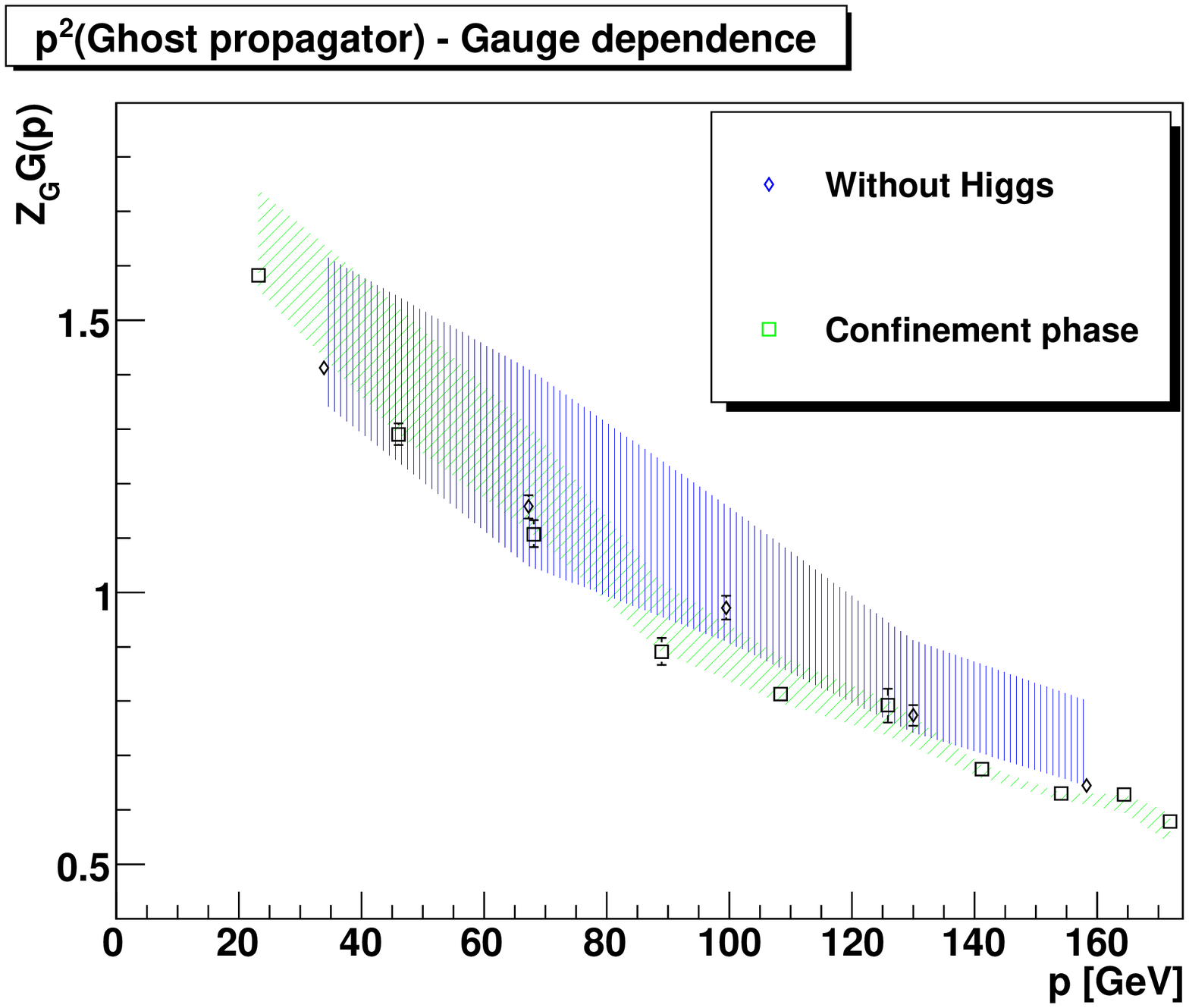}\\
\includegraphics[width=0.5\linewidth]{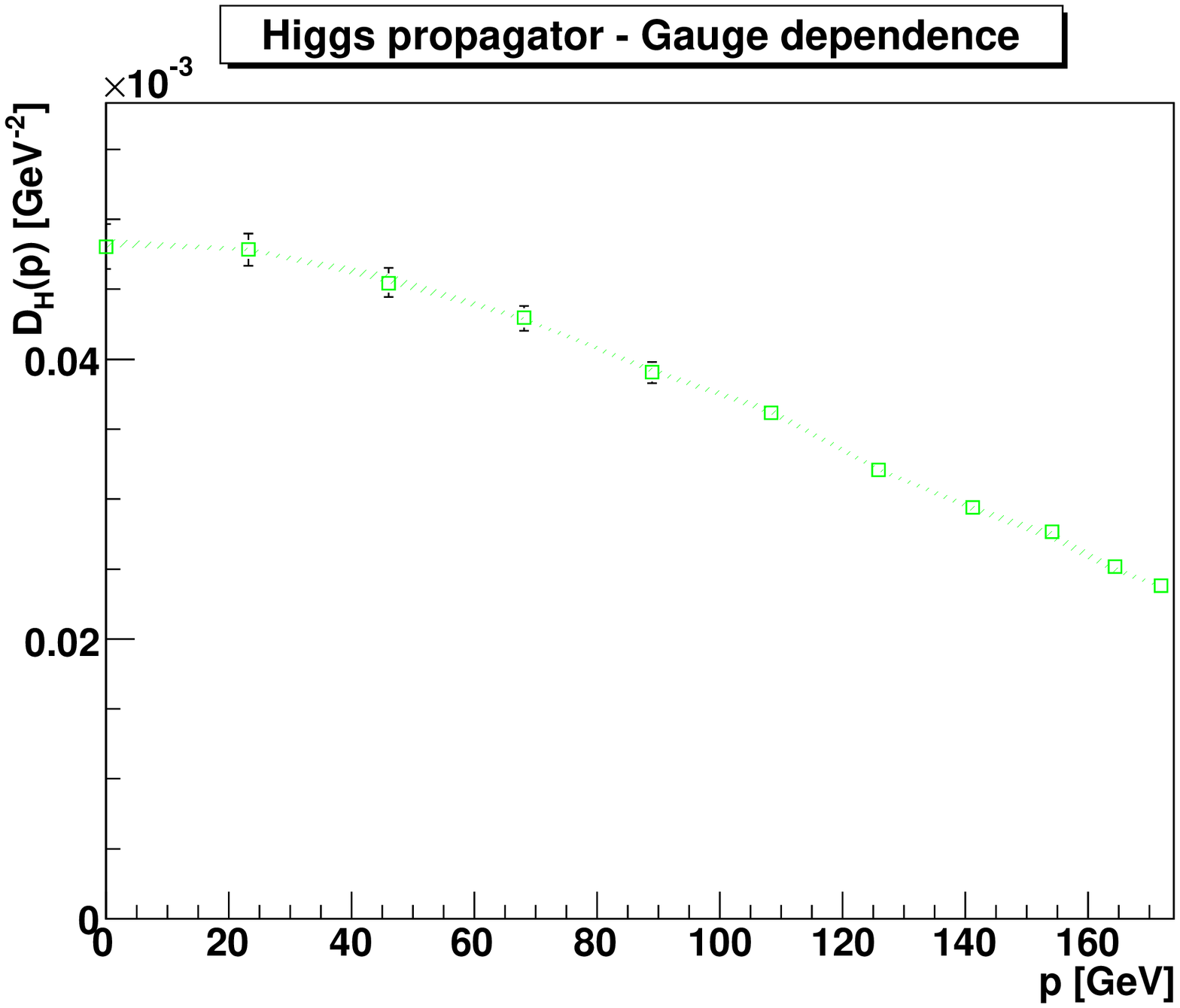}\includegraphics[width=0.5\linewidth]{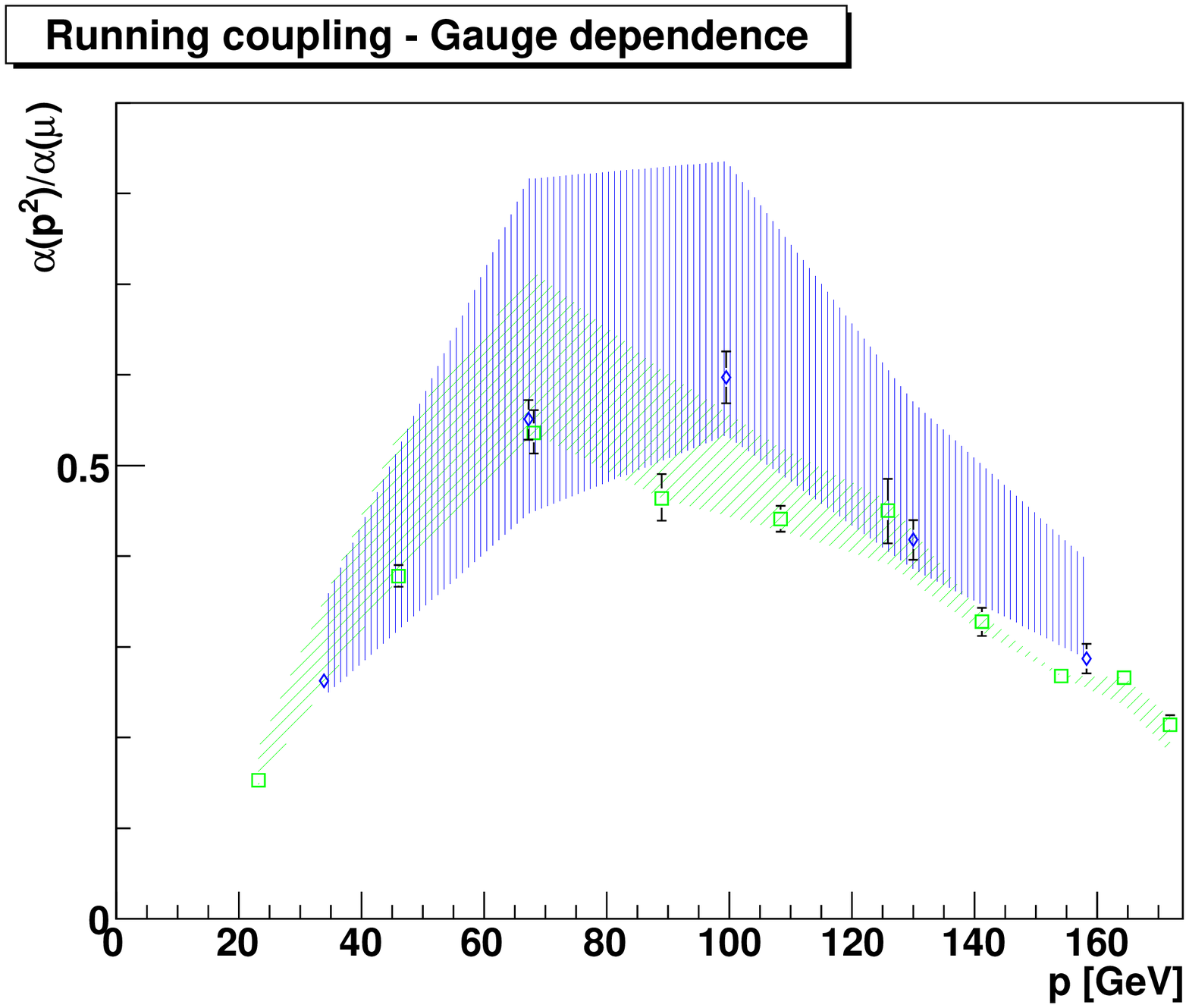}
\caption{\label{fig:gribov}The gauge dependence of the correlation functions in comparison between the Yang-Mills theory and the confinement phase. The top left panel shows the $W$ propagator, the top right panel the ghost dressing function, the bottom-left panel the Higgs propagator (only the confinement phase is shown, see text), and the bottom right panel shows the running gauge coupling. The data points are in the minimal Landau gauge, and the hatched area represents the variability of the average value with the gauge. Diagonally hatched areas are for the confinement phase, and vertically hatched areas are for the case of Yang-Mills theory. For clarity, scale uncertainties have been suppressed in this figure, as they are not relevant since the ratio of scales of both cases is fixed.}
\end{figure*}

The result for the three correlation functions and the running coupling are shown in figure \ref{fig:gribov}. The dependence of the quenched Higgs propagator is left out intentionally, since it does not represent a dynamical variable, and therefore its gauge dependence is irrelevant for the present purpose.

For the ghost propagator the gauge variability exceeds the statistical error. Thus, it is significantly gauge-de\-pen\-dent, even given the small numbers of Gribov copies used. In fact, its dependence is expected to increase significantly even further when enlarging the search space \cite{Silva:2004bv,Maas:2009ph}, as is the case in Yang-Mills theory, if the ratio of search space to average number of Gribov copies found is so close to one as in the present case. On the other hand, the $W$ propagator shows no change within the statistical error, though when including more Gribov copies this is known to change for pure Yang-Mills theory at fixed lattice parameters \cite{Bornyakov:2008yx}. Nonetheless, the running coupling, derived from the gauge propagators, is consequently significantly dependent on the gauge. In contrast, the Higgs propagator is not gauge-dependent within the statistical errors. This is not surprising, as the gauge conditions include only the ghost and $W$ fields. However, if a 't Hooft gauge outside the Landau limit is used, this may change. Still, the gauge-dependence in the case with Higgs is rather similar to that in the Yang-Mills case for all correlation functions, again emphasizing the similarity between both cases.

In total, the non-perturbative gauge-dependence in the confinement phase is sizable, and should be kept in mind if, e.\ g., results should be transferred between different calculations or even methods. On the other hand, within the limited scope of this investigation, no effect in the Higgs phase is found.

\section{Summary}\label{ssum}

Summarizing, the present study shows the feasibility of determining the gauge-dependent correlation functions in Yang-Mills-Higgs systems beyond perturbation theory, in both the confinement and the Higgs phase, even for a very heavy Higgs. It therefore provides a route to unify the non-perturbative gauge-invariant description usually employed in lattice calculations with the approach of perturbation theory, which operates directly on the gauge-dependent elementary degrees of freedom. In addition, first hints how to provide such a relation have been obtained.

This exploratory investigation has furthermore shown that the difference between the would-be confinement phase and the would-be Higgs phase is essentially manifest in the gauge-fixing sector outside the physical domain, in agreement with previous studies and arguments \cite{Greensite:2004ur,Kugo:1979gm,Alkofer:2000wg}. It has also been shown that, within the limited range of gauges investigated here, there is little qualitative difference between both phases otherwise. However, the behavior found in the would-be Higgs phase is in agreement with the expectations for a dynamically induced Higgs effect. On the other hand, the confinement phase showed little difference compared to Yang-Mills theory. This is similar to the case when including dynamical fundamental quarks instead of scalars, where the difference haven been found to be also rather small (see, e.\ g.,\ \cite{Bowman:2007du,Fischer:2006ub}).

In the present case, it is impossible to assess what the continuum and infinite-volume limit of these results will be, if the theory should be non-trivial. Therefore, it must now be the primary aim to further develop this approach, and push in both phases towards the continuum field theory, or at least to a useful cutoff-theory in case of triviality. This will not be simple in this approach, since the Higgs propagator itself will not be sufficient to differentiate between an only very weakly interacting theory and a trivial one. This requires determination of the vertices, a technique well-developed for Yang-Mills theory \cite{Cucchieri:2006tf,Cucchieri:2008qm}, and straightforwardly applicable to the present setting. Also, results in the confinement phase are hopefully shedding light on how the confinement of fundamental charges in dynamical theories is manifest in low-order correlation functions \cite{Fister:2010yw}, and how it is different from screening due to the Higgs effect.\\

{\bf\no Acknowledgments}

I am grateful to C.\ B.\ Lang and O.\ Rosten for helpful discussions and to R.\ Alkofer, J.\ Greensite, and J.\ M.\ Pawlowski for comments on the manuscript. This work was supported by the FWF under grant number M1099-N16. The ROOT framework \cite{Brun:1997pa} has been used in this project. Computing time was provided by the HPC center at the Karl-Franzens-University Graz and the Slovak Grant Agency for Science, Grant VEGA No.\ 2/6068/2006.

\appendix

\section{Determination of the scale and systematic uncertainty}\label{app:mass}

\begin{figure}
\includegraphics[width=\linewidth]{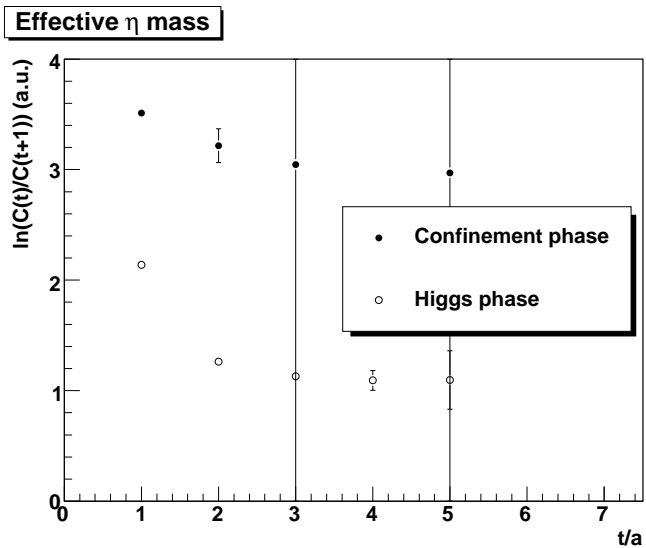}
\caption{\label{hhmass}The effective mass \cite{Gattringer:2010zz} as a function of time.}
\end{figure}

\begin{figure}
\includegraphics[width=\linewidth]{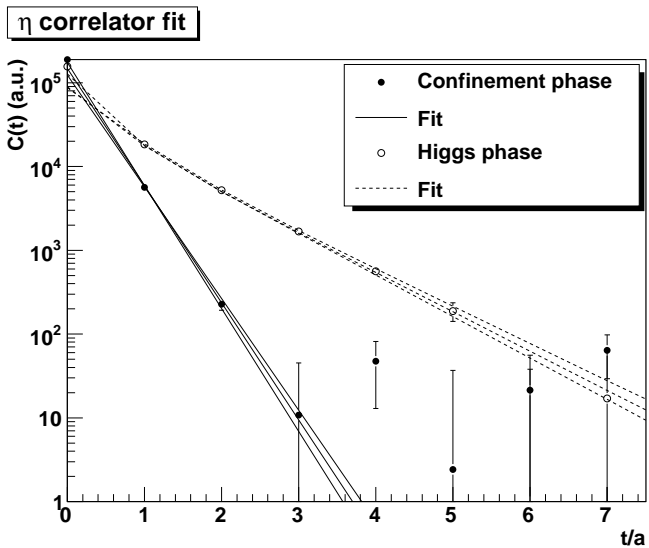}
\caption{\label{hh}The quality of determination of the scale. Shown is the raw data together with a fit of the type \pref{coshfit}.}
\end{figure}

\begin{table*}
\caption{\label{hhtab}The results for the fit of type \pref{coshfit}. For the fits, the number of configurations indicated have been used, which have been generated in multiple runs with 1080 thermalization sweeps and 24 decorrelation sweeps.}
\vspace{1mm}
\begin{tabular}{|c|c|c|c|c|c|}
\hline
System & Configurations & a & b & m & n \cr
\hline
Confinement & 1265277 & 49$^{+198}_{-40}\times 10^{-13}$ & - & 3.2(2) & - \cr
\hline
Higgs & 471146 & [9$\times 10^{-14}$,5$\times 10^{-7}$] & 19$^{+15}_{-8}\times 10^{-2}$ & $2.6_{-0.4}^{+0.9}$ & $1.08^{+0.06}_{-0.04}$ \cr
\hline
\end{tabular}
\end{table*}

As noted in section \ref{stech}, the ground-state energy of the state \pref{higgsonium} is used to set the scale. This energy can be extracted from the exponential decay of its correlation function $C(t)$ at asymptotically large times \cite{Gattringer:2010zz}, in particular from a plateau in $\ln(C(t+1)/C(t))$. However, the correlator may also contain contributions from excited states at finite times. As is visible in figure \ref{hhmass}, this is indeed the case, at least in the Higgs phase. Thus, it is necessary to take the presence of excited states explicitly into account. A convenient way is to fit the correlator with the ansatz \cite{Gattringer:2010zz}
\be
C_f(t)=a\cosh(m(t-N/2))+b\cosh(n(t-N/2))\label{coshfit},
\ee
\no where all quantities are in lattice units. The result, together with the raw data, is shown in figure \ref{hh}. Errors are determined by fitting besides the central value the one-$\sigma$ intervals, and as fit intervals always the time interval is used for which the correlation function is a positive, monotonously decaying function of time, and is adjusted for both upper and lower one-$\sigma$ bounds.

Despite the large number of configurations included, see table \ref{hhtab}, it is in particular for the confinement case not possible to unambiguously identify an excited state. Usage of an enlarged base of operators may here be useful in future calculations \cite{Gattringer:2010zz}. Therefore, only a single state fit is possible. The resulting fit parameters, together with the number of configurations, are given in table \ref{hhtab}. The procedure in the main text is then used to associate a scale with the lower mass in the Higgs phase and the only mass in the confinement phase. The resulting value is given in table \ref{config}. The error on the scale is just the statistical error. It is propagated through all calculations in section \ref{sprop}. However, the dressing functions, and thus also the running coupling depending only on these, are dimensionless quantities, and therefore only the determination of the momentum in these cases receive an additional error. Furthermore, since the quenched case has a uniquely assigned scale based on the scale in the confinement case, the scale error is irrelevant when comparing only the confinement and the quenched case, and can therefore be dropped in section \ref{sgauge}.

\begin{figure}
\includegraphics[width=\linewidth]{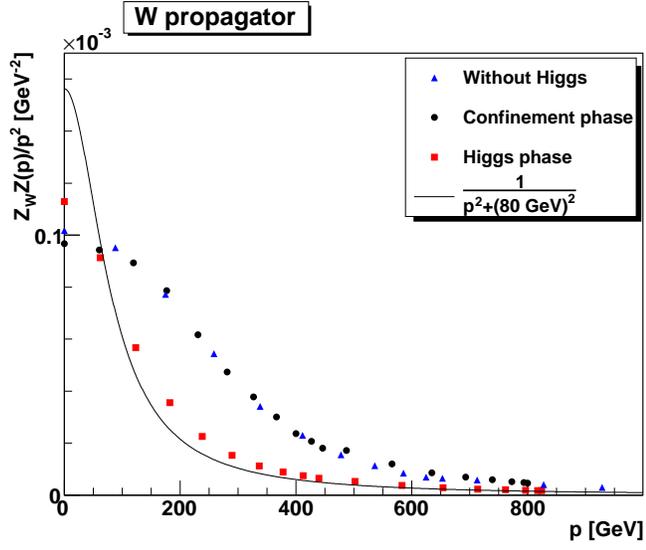}\\
\includegraphics[width=\linewidth]{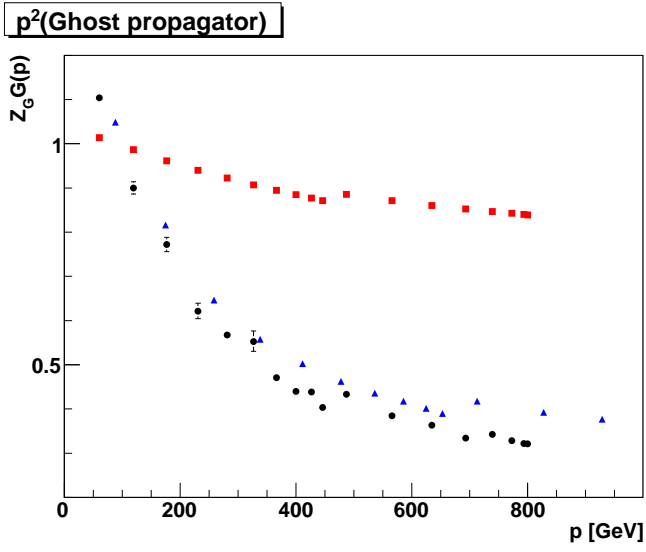}\\
\includegraphics[width=\linewidth]{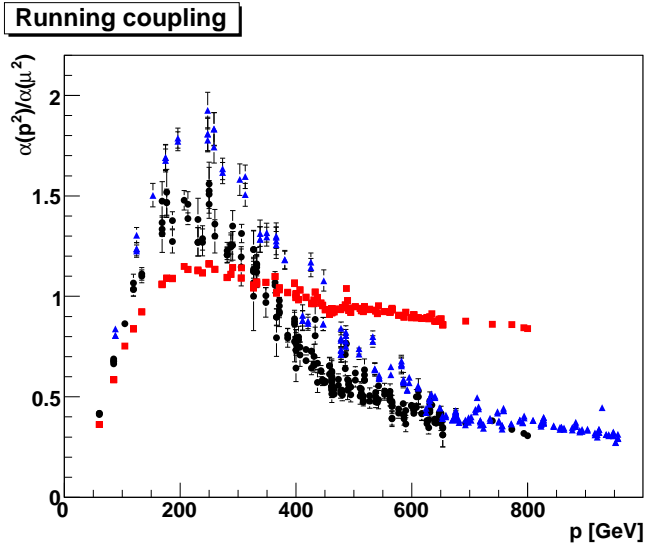}
\caption{\label{sys-g}The $W$ propagator (top panel), the ghost dressing function (middle panel), and the running coupling (bottom panel) under the assumption of equal scales for the Higgs and the confinement case.}
\end{figure}

\begin{figure}
\includegraphics[width=\linewidth]{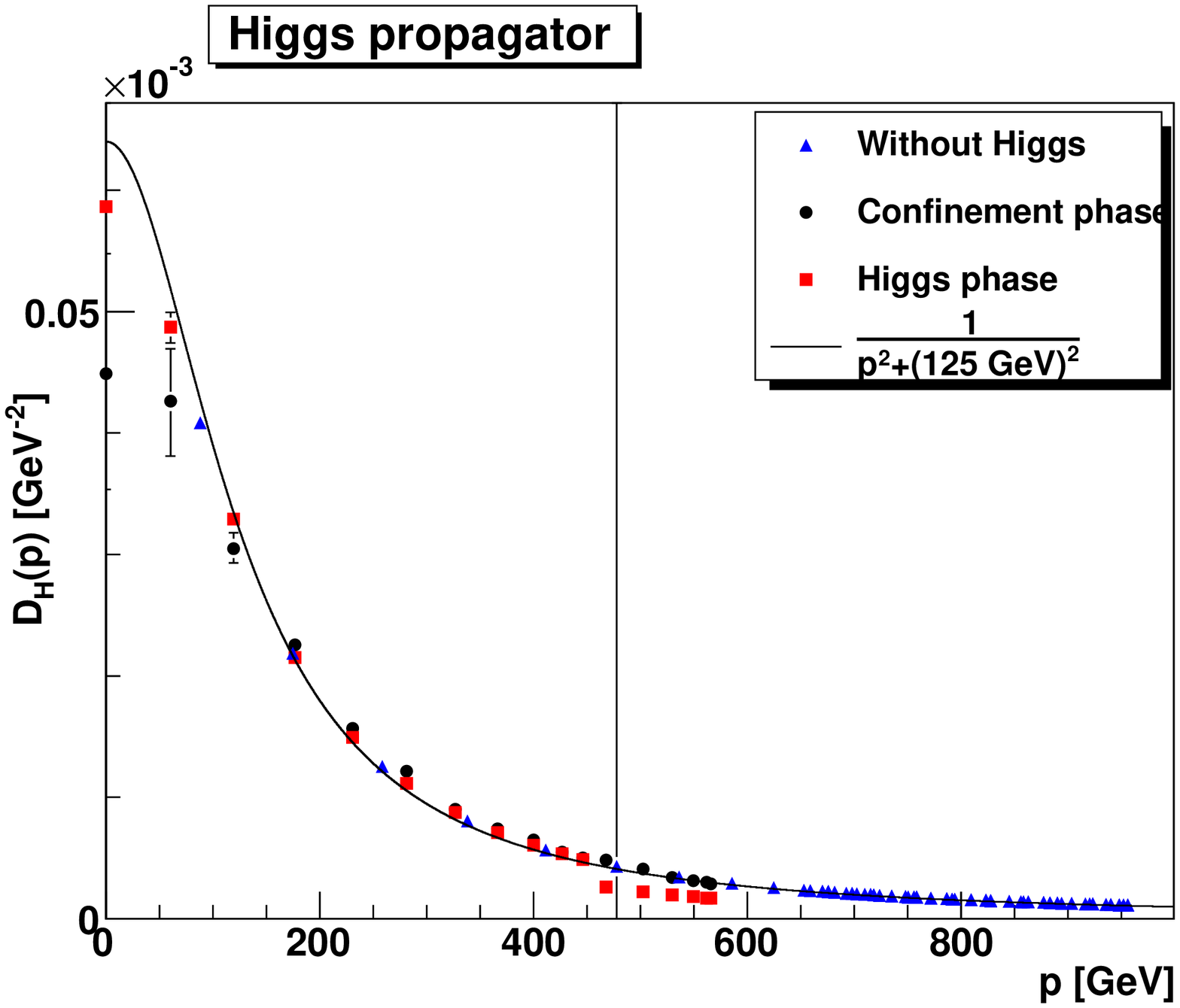}
\caption{\label{sys-s}The Higgs propagator under the assumption of equal scales for the Higgs and the confinement case. the single large error bar is due to a numerical coincidence in the error propagation of the renormalization process, due to the appearance of a derivative.}
\end{figure}

Unfortunately, with the available lattice techniques it can never be guaranteed that there is not a further state still present, which is still lighter than the lightest one found, in particular if having a very small pre-factor. This is a source of systematic uncertainty in the scale determination. To illustrate this uncertainty, it will now be assumed that in the confinement phase there exists a lighter state with the same mass as the lighter state in the Higgs phase. For this hypothetical situation the error on this scale can be ignored. The results for the $W$ sector in this case are shown in figure \ref{sys-g} and for the Higgs sector in figure \ref{sys-s}. In the gauge sector nothing dramatic changes. Only the $W$ propagator becomes less tree-level-like, while the running coupling becomes somewhat more similar between both cases. Still, the ghost is the one most different between both cases, being a strongly momentum-dependent function in the confinement phase, but almost momentum-independent in the Higgs case. In the Higgs case, the changes are essentially negligible. Thus, the statements made in the main text are likely only (weakly) quantitatively affected by the systematic uncertainty in the scale determination.

\bibliographystyle{bibstyle}
\bibliography{bib}


\end{document}